\documentclass[a4paper,11pt]{article}

\usepackage{amsmath}
\usepackage{amssymb}
\usepackage[applemac]{inputenc}
\usepackage{latexsym}
\usepackage{graphicx}
\usepackage{color}
\usepackage[german, frenchb, english]{babel}

\usepackage{amsfonts}
\input cyracc.def

\newcommand{\id}{{\rm id}}
\newcommand{\ad}{{\rm ad}}

\newcommand{\ie}{{\rm i}}

\newcommand{\Tr}{{\rm Tr}\,}
\newcommand{\del}{\partial}

\newcommand{\Hc}{{\mathcal H}}

\newtheorem{thm}{Theorem}[section]
\newtheorem{prop}[thm]{Proposition}
\newtheorem{rem}[thm]{Remark}
\newtheorem{lem}[thm]{Lemma}
\newtheorem{cor}[thm]{Corollary}

\newtheorem{defe}[thm]{Definition}

\def\proofp{{\hspace{-0.5 cm} $\Box$ \textbf{Proof of Proposition }}}
\def\proofc{{\hspace{-0.5 cm} $\Box$ \textbf{Proof of Corollary
      }}}
\def\prooft{{\hspace{-0.5 cm} $\blacksquare$
\textbf{Proof of Theorem }}}
\def\proofl{{\hspace{-0.5 cm} $\vartriangle$ \textbf{Proof of Lemma  }}}
\def\dem{{\hspace{-0.5 cm} $\Box$ \textbf{Proof of Proposition }}}
\def\demcor{{\hspace{-0.5 cm} $\Box$ \textbf{Proof of Corollary  }}}
\def\demthm{{\hspace{-0.5 cm} $\blacksquare$
\textbf{Proof of Theorem  }}}
\def\demlem{{\hspace{-0.5 cm} $\vartriangle$ \textbf{Proof of Lemma }}}

\def\cqfd{{\hfill $\Box$}}
\def\cqfdt{{\hfill $\blacksquare$}}
\def\cq{{\hfill $\vartriangle$}}

\def\C{{\mathbb{C}}}
\def\R{{\mathbb{R}}}

\def\N{{\mathbb{N}}}

\newcommand{\gm}{{\rm g}}

\def\g{{\mathfrak{g}}}
\def\ha{{\mathfrak{h}}}
\def\k{{\mathfrak{k}}}

\def\a{{\mathfrak{a}}}
\def\b{{\mathfrak{b}}}

\begin{document}
\title{ Mostow's Decomposition Theorem for $L^{*}$-groups and
Applications to  affine coadjoint orbits and stable manifolds}

\author{Alice Barbara TUMPACH\footnote{{\tt alice.tumpach@epfl.ch}, EPFL, Lausanne, Switzerland.
   This
  work was partially supported by
  the University of Paris VII, the University of
  Paris XI, and the \'Ecole Polytechnique, Palaiseau, France.}}

\date{}

\maketitle

\abstract Mostow's Decomposition Theorem is a refinement of the
polar decomposition. It states the following. Let $G$ be a compact
connected semi-simple Lie group with Lie algebra $\g$. Given a
subspace $\ha$ of $\g$ such that $[X, [X, Y]] \in \ha$ for all
$X$, $Y$ in $\ha$, the complexified group $G^{\C}$ is homeomorphic
to the product $G \cdot \exp \ie\mathfrak{m} \cdot \exp \ie\ha$,
where $\mathfrak{m}$ is the orthogonal of $\ha$ in $\g$ with
respect to the Killing form. This Theorem is related to geometric
properties of the non-positively curved space of positive-definite
symmetric matrices and to a characterization of its geodesic
subspaces. The original proof of this Theorem given by Mostow in
\cite{Mostow} uses the compactness of $G$. We give a
 proof of this Theorem using the completeness of the Lie
algebra $\g$ instead, which can therefore be applied to an
$L^*$-group of arbitrary dimension. A different proof of this
generalization of Mostow's Decomposition Theorem has been obtained
independently by G.~Larotonda in \cite{Lar}. Some applications of
this Theorem to the geometry of  (affine) coadjoint orbits and
stable manifolds are given.

\begin{center}
\textbf{R\'esum\'e}
\end{center}
Le th\'eor\`eme de d\'ecompo\-si\-tion de Mostow est un
raffinement de la d\'ecomposition polaire. Il s'\'enonce comme
suit. Soit $G$ un groupe de Lie compact connexe semi-simple
d'alg\`ebre de Lie $\g$. \'Etant donn\'e un sous-espace vectoriel
$\ha$ de $\g$ tel que $\left[X, [X, Y]\right] \in \ha$ pour tout
$X$, $Y$ dans $\ha$, le groupe de Lie complexifi\'e $G^{\C}$ de
$G$ est hom\'eomorphe au produit $G \cdot \exp \ie\mathfrak{m}
\cdot \exp \ie\ha$, o\`u $\mathfrak{m}$ est l'orthogonal de $\ha$
dans $\g$ relativement \`a la forme de Killing. Ce th\'eor\`eme
repose sur la g\'eom\'etrie \`a courbure n\'egative de l'espace
des matrices sym\'etriques d\'efinies positives, et sur la
caract\'erisation de ses sous-espaces totalement g\'eod\'esiques.
La preuve initiale de ce th\'eor\`eme donn\'ee par Mostow dans
\cite{Mostow} utilise la compacit\'e de $G$. Nous en donnons une
d\'emonstration qui utilise seulement la compl\'etude de
l'alg\`ebre de Lie $\g$, et qui s'applique de ce fait \`a un
$L^*$-groupe de dimension arbitraire. Une autre d\'emonstration de
cette g\'en\'eralisation du th\'eor\`eme de d\'ecomposition de
Mostow a \'et\'e obtenue ind\'ependamment par G.~Larotonda dans
\cite{Lar}. Quelques applications de ce th\'eor\`eme \`a la
g\'eom\'etrie des orbites coadjointes (affines) et des surfaces
stables sont donn\'ees.

 \tableofcontents

\section{Mostow's Decomposition Theorem for $L^{*}$-groups}

\subsection{Introduction}

This section is devoted to a  proof of Mostow's Decomposition
Theorem for a separable $L^*$-group of finite or infinite
dimension. Some general results in the geometry of Banach
manifolds are stated passing by, but have their interest on their
own (for instance Proposition \ref{genial}). We refer to
\cite{Mostow} for the original arguments in the finite-dimensional
case, to \cite{Lar} for a different proof in the
infinite-dimensional setting and to \cite{AL} for a generalization
to some von Neumann algebras. Let us first state the Theorem~:

\begin{thm}\label{debut}
 Let $G$ be a semi-simple connected
$L^{*}$-group of compact type with Lie algebra  $\mathfrak{g}$,
$G^{\C}$ the connected $L^{*}$-group with Lie algebra
$\mathfrak{g}^{\C} := \mathfrak{g} \oplus \ie \g$, $E$ a closed
subspace of  $~\ie\g$ such that
$$
\left[ X, [X, Y]\right] \in E, \qquad for~all~~~ X, Y \in E,
$$
and $F$ the orthogonal of $E$ in $\ie\g$. Then $G^{\C}$ is
homeomorphic to the product  $G \cdot \exp F \cdot \exp E$.
\end{thm}

This section is organized as follows. First we investigate the
geometry of the space $\mathcal{P}$ of positive-definite
self-adjoint operators in the group $\textrm{GL}_{2}(\Hc)$ of
invertible operators which differ from the identity by
Hilbert-Schmidt operators. We show in particular that it is a
symmetric space of non-positive sectional curvature and that the
exponential map defined by the usual power series is  a
diffeomorphism from the space $\mathcal{S}_{2}(\Hc)$ of
Hilbert-Schmidt self-adjoint operators onto $\mathcal{P}$.
Moreover we show that the exponential map is the Riemannian
exponential map at the identity with respect to the defined metric
on $\mathcal{P}$. This is implied by a general result on the
geodesics in locally symmetric spaces that we state in the more
general context of Banach manifolds (Proposition \ref{genial}).
This study implies the usual Al-Kashi inequality on the sides of a
geodesic triangle in the non-positively curved space
$\mathcal{P}$, and the convexity property of the distance between
two geodesics. In the second subsection, a characterization of the
geodesic subspaces of $\mathcal{P}$ is given which mainely follows
\cite{Mostow}. In subsection \ref{sub3}, the key-step for the
proof of Mostow's Decomposition Theorem is given by the
construction of a projection from $\mathcal{P}$ to every closed
geodesic subspace. The arguments given here for the existence of
such projection are simpler and more direct then the ones given in
the original paper \cite{Mostow}, and apply to arbitrary
dimension. In the last subsection, we use this projection to prove
the Theorem stated above. For examples of application of this
Theorem we refer to sections \ref{section2} and \ref{section3} of
the present work.

\subsection{The space $\mathcal{P}$ of positive-definite self-adjoint
  operators in $\textrm{GL}_{2}(\Hc)$}

Let $\Hc$ be an Hilbert space of arbitrary dimension. The group
$\textrm{GL}_{2}(\Hc)$ is the group of invertible operators which
differs from the identity by an Hilbert-Schmidt operator~:
$$
\begin{array}{l}
\textrm{GL}_{2}(\Hc) := \{g\in\textrm{GL}(\Hc)\mid a-\id\in
L^{2}(\Hc)\},
\end{array}
$$
where $L^2(\Hc)$ denotes the Hilbert space of Hilbert-Schmidt
operators on $\Hc$, i.e. the set of operators $A$ on $\Hc$ such
that the trace of $A^*A$ is finite, endowed with the Hermitian
scalar product~:
$$
\langle A, B \rangle := \Tr A^*B.
$$
The Lie-algebra of $\textrm{GL}_{2}(\Hc)$ is the Hilbert space
 $L^2(\Hc)$ endowed with the commutator of operators, and will be
denoted by $\mathfrak{gl}_{2}(\Hc)$. It is an $L^*$-algebra for
the involution $*$ which maps an operator to its adjoint, in the
sense that the bracket on $\mathfrak{gl}_{2}(\Hc)$ and the
involution $*$ are related by the following property~:
$$ \langle [x\,,\,y]\,,\,z\rangle =
\langle y\,,\,[x^{*},\,z]\rangle $$ for every $x$, $y$ and $z$ in
$\mathfrak{gl}_{2}(\Hc)$. We define also the unitary group
$\textrm{U}_{2}(\Hc)$ and its Lie algebra
$\mathfrak{u}_{2}(\Hc)$~:
$$
\begin{array}{l}
\textrm{U}_{2} = \{a\in\textrm{U}(\Hc)\mid a-\id\in L^2(\Hc)\},\\
\mathfrak{u}_{2}=\mathfrak{u}(\Hc)\cap L^2(\Hc).
\end{array}
$$
The Hilbert space $\mathfrak{gl}_{2}(\Hc)$ splits into the direct
sum of $\mathfrak{u}_{2}(\Hc)$ and the closed linear subspace
$\mathcal{S}_{2}(\Hc)$ of self-adjoint elements in
$\mathfrak{gl}_{2}(\Hc)$. The exponential map defined as
\begin{equation}\label{exp}
\exp(A) := \sum_{n = 0}^{+ \infty} \frac{A^{n}}{n!}
\end{equation}
for all $A$ in $\mathfrak{gl}_{2}(\Hc)$, takes
$\mathcal{S}_{2}(\Hc)$ to the submanifold $\mathcal{P}$ of
$\textrm{GL}_{2}(\Hc)$ consisting of positive-definite
self-adjoint operators~:
$$
\exp~: \mathcal{S}_{2}(\Hc) = \{ A \in \mathfrak{gl}_{2}(\Hc),
A^{*} = A \} \longrightarrow \mathcal{P} = \{ A \in
\mathcal{S}_{2}(\Hc), A^{*}A
> 0 \}.
$$
Note that for $\a \in \mathfrak{gl}_{2}(\Hc)$, the curve
$\gamma(t):= \exp (t\a)$ satisfies $\dot{\gamma}(t) =
\left(L_{\gamma(t)}\right)_{*}(\a)$, where $L_{\gamma(t)}$ denotes
left translation by $\gamma(t)$, hence the exponential map defined
by \eqref{exp} is the usual exponential map on the Lie group
$\textrm{GL}_{2}(\Hc)$.
 Note also that the tangent space to
$\mathcal{P}$ at $p~$ is obtained from $\mathcal{S}_{2}(\Hc)$ by
left or right translation by $p$. Let us endowed $\mathcal{P}$
with the following Riemannian metric~:
$$
\gm_{{p}}(U, V) = \Tr\left(p^{-1}U p^{-1}V\right).
$$
Note that for $U, V \in T_{p}\mathcal{P}$, $p^{-1}U$ and $p^{-1}V$
belong to $\mathcal{S}_{2}(\Hc)$, thus the product $p^{-1}U
p^{-1}V$ is of trace class, so $\gm$ is well-defined and
positive-definite. The identity on $\Hc$ belongs to $\mathcal{P}$
and as such will be denoted by $o$ in order to avoid confusions
with identity operators on other Hilbert spaces.

\begin{prop}\label{isom}
The action  of $\textrm{GL}_{2}(\Hc)$ on $\mathcal{P}$ defined
by~:
$$
\begin{array}{rll}
\textrm{GL}_{2}(\Hc) \times \mathcal{P} &\rightarrow& \mathcal{P}\\
 (\, x\,,\,p\, ) & \mapsto & x\cdot p = x^{*} p \,\,x,
\end{array}
$$
is a transitive action by isometries.
\end{prop}

\dem \ref{isom}:\\
For every $p$ in $\mathcal{P}$, the square root of $p$ is
well-defined and belongs to $\mathcal{P}$. In other words there
exists $q$ in $\mathcal{P}$ such that $p = q^{2}$. But $q^* = q$,
hence $p = q^{*} q= q\cdot o$, and the transitivity follows. To
show that $\textrm{GL}_{2}(\Hc)$ acts by isometries, for $x$ in
$\textrm{GL}_{2}(\Hc)$ and $p \in \mathcal{P}$, let us denote by
$x_*$ the differential of $x$ at $p$. By linearity, one has
$x_{*}(U) = x^{*} U x$ for every $U\in T_{p}\mathcal{P}$. It
follows that
$$
\begin{array}{ll}
\gm_{x\cdot p}\left(x_*(U)\,,\,x_*(V)\right) & =  \Tr (x\cdot
p)^{-1}x_*(U) (x\cdot p)^{-1}x_{*}(V) = \Tr x^{-1} p^{-1}
(x^{*})^{-1}x^{*} U x x^{-1} p^{-1} (x^{*})^{-1} x^{*} V x \\
& =  \Tr x^{-1} p^{-1} U p^{-1} V x = \Tr p^{-1} U p^{-1}V =
\gm_{p}(U, V),
\end{array}
$$
and the action of $\textrm{GL}_{2}(\Hc)$ on $\mathcal{P}$ is an
action by isometries.\cqfd

\begin{rem}
{\rm Since the stabilizer of the identity is the unitary group
$\textrm{U}_{2}(\Hc)$, it follows that\\
$\mathcal{P}~=~\textrm{GL}_{2}(\Hc)/U_{2}(\Hc)$ as homogeneous
Riemannian manifold.}
\end{rem}

Let us recall the following definition.
\begin{defe}{\rm
A Riemannian Banach manifold $\mathcal{P}$ is called
\emph{symmetric} if for every $p$ in $\mathcal{P}$, there exists a
globally defined isometry $s_{p}$ which fixes $p$ and such that
the differential of $s_{p}$ at $p$ is $-\id$.}
\end{defe}

\begin{prop}\label{symneg}
The manifold $\mathcal{P}$ is a symmetric homogeneous Riemannian
manifold of non-positive sectional curvature.
\end{prop}

\dem \ref{symneg}:\\
Consider the inversion $i\,: \mathcal{P} \rightarrow \mathcal{P}$
defined by $i(p) = p^{-1}$. An easy computation shows that the
differential $i_{*}$ of $i$ at $p \in \mathcal{P}$ takes each
vector $U \in T_{p}\mathcal{P}$ to $i_{*}(U) = -p^{-1} U p^{-1}$.
One therefore has~:
$$
\begin{array}{ll}
\gm_{i(p)}\left( i_{*}(U), i_{*}(V) \right) & = \gm_{p^{-1}}\left(
-p^{-1} U p^{-1}, -p^{-1} V p^{-1}\right) = \Tr p(-p^{-1} U
p^{-1}) p (-p^{-1} V p^{-1})\\
& = \Tr U p^{-1} V p^{-1} = \Tr p^{-1} U p^{-1} V = \gm_{p}(U, V),
\end{array}
$$
hence $i$ is an isometry of $\mathcal{P}$. Since $i$ fixes the
 element $e = \id_{\Hc}$ and since the differential of $i$ at $e$ is
$-\id$ on $T_{e}\mathcal{P} = \mathcal{S}_{2}(\Hc)$, it follows
that $i$ is a global symmetry with respect to $e$. It follows from
the transitive action of $\textrm{GL}_{2}(\Hc)$ that for every $p$
in $\mathcal{P}$ there exists a global isometry $s_{p}$ which
fixes $p$ and whose differential at $p~$ is $-\id$ on
$T_{p}\mathcal{P}$, namely $s_{p}(x) = p~x^{-1}\,p$. Whence
$\mathcal{P}$ is a symmetric homogeneous Riemannian manifold.

Since the $\textrm{GL}_{2}(\Hc)$-invariant metric on $\mathcal{P}$
is a \emph{strong} Riemannian metric (this means that $\gm$
induces an isomorphism between the tangent space at $p \in
\mathcal{P}$ and its continuous dual, which is clearly the case at
$e$ hence everywhere), the Levi-Civita connection is well-defined
by Koszul formula (see for instance Theorem 3.1 page~54 in
\cite{Ar03} where this formula is recalled). The computation of
the curvature tensor $R$ is therefore step by step the same as in
the finite-dimensional case and provides that the sectional
curvature $K_{o}$ at $o$ is given by~:
$$
K_{o}(X, Y) := \frac{\gm\left( R_{X, Y}X, Y\right)}{\gm(X,
X)\gm(Y, Y) -\gm(X, Y)^{2}} = \frac{\gm\left(\left[ [X, Y],
X\right], Y \right)}{\gm(X, X)\gm(Y, Y) -\gm(X, Y)^{2}},
$$
for all $X$, $Y$ in $T_{o}\mathcal{P} = \mathcal{S}_{2}(\Hc)$ (see
Proposition 7.72 page~193 in \cite{Bes}, or Proposition 6.5
page~92 in \cite{Ar03}). Now the sign of the sectional curvature
of the $2$-plane generated by $X$ and $Y$ is the sign of
$\gm\left(\left[ [X, Y], X\right], Y \right)$. By definition of
$\gm$, one has~:
$$
\begin{array}{ll}
\gm\left(\left[ [X, Y], X\right], Y \right) & = \Tr \left[ [X, Y],
X\right] Y = \Tr \left([X, Y] X Y - X [X, Y] Y \right)\\& = \Tr
[X, Y] [X, Y] = - \Tr [X, Y]^{*} [X, Y] \leq 0,
\end{array}
$$
the last identity following from the fact that  $[X, Y]$ belongs
to $[\mathcal{S}_{2}(\Hc), \mathcal{S}_{2}(\Hc)] \subset
\mathfrak{u}_{2}(\Hc)$.
 \cqfd
\\
\\

The following Proposition is well-known in the theory of linear
group. The reader will find the computation of the differential of
the exponential using powers series as a consequence of Lemma 1 in
\cite{Mostow} (this computation works as well in the
infinite-dimensional setting, see for instance Proposition 2.5.3
page 116 in \cite{Tum}). See also \cite{God}.

\begin{prop}\label{log}
For every Hilbert Lie group $G$, with Lie algebra $\g$, the
differential of the exponential map $~\exp~: \g \rightarrow G$ is
given at $X \in \g$ by~:
\begin{equation}\label{difexp}
\left(d_{X}\exp\right)(Y) = L_{\exp(X)}\left(\frac{1 -
  e^{-\ad(X)}}{\ad(X)}\right)(Y).
\end{equation}
for all $Y$ in $\g$.
\end{prop}

\dem \ref{log}:\\
Let us define the following map~:
$$
\begin{array}{llll}
\Phi~: & \mathbb{R}^2 & \longrightarrow & G\\
       & (t, s) & \longmapsto & \exp\left(t(X + sY)\right).
\end{array}
$$
Consider the push-forward $U$ and $V$ of the vector fields
$\frac{\del}{\del t}$ and $\frac{\del}{\del s}$ on
$\mathbb{R}^2$~:
$$U\left(\Phi(t, s)\right) := \Phi_{*}\left(\frac{\del}{\del t}\right)~~\textrm{and}~~ V\left(\Phi(t,
s)\right) = \Phi_{*}\left(\frac{\del}{\del s}\right).$$ Denote by
$[\cdot\,,\,\cdot]_{\mathfrak{X}}$ the bracket of vector fields.
One has:
\begin{equation}\label{comUV}
[U, V]_{\mathfrak{X}} = \left[\Phi_{*}\left(\frac{\del}{\del
t}\right), \Phi_{*}\left(\frac{\del}{\del
s}\right)\right]_{\mathfrak{X}} = \Phi_{*}\left[\frac{\del}{\del
t}, \frac{\del}{\del s}\right]_{\mathfrak{X}} = 0.
\end{equation}
 Note that
$$
V\left(\Phi(t, s)\right) = \frac{\del \Phi}{\del s}(t, s) =
\left(d\exp_{(tX + stY)}\right)(tb) ~~\textrm{and}~~V\left(\Phi(1,
0)\right) = \left(d_{X}\exp\right)(Y).
$$
The idea of this proof is to explicit the differential equation
satisfied by the $\g$-valued function
$${v}(t):=
\left(L_{\Phi(t, 0)}^{-1}\right)_{*}V\left(\Phi(t, 0)\right) =
\left(L_{\exp(tX)}^{-1}\right)_{*}V\left(\exp(tX)\right).$$
 For this purpose
let us introduce the connection $\nabla$ on the tangent bundle of
$G$ for which the left-invariant vector fields are parallel. It is
a flat connection since a trivialization of the tangent bundle is
given by the left-invariant vector fields associated to an Hilbert
basis of $\g$. Note that by the very definition of the exponential
map on a Lie group, $t \mapsto \exp(tX) = \Phi(t, 0)$ is a
geodesic for this connection, hence
\begin{equation}\label{unablau}
\nabla_{U}U = 0
\end{equation}
 along $\Phi(t, 0)$. More generally,  the connection $\nabla$ can be expressed
 using the (left) Maurer-Cartan $\g$-valued $1$-form
defined by
$$
\theta_{g}(Z) = (L_{g})^{-1}_{*}(Z).
$$
Indeed, for a vector field $W$ and a tangent vector $Z$ in
$T_{g}G$, one has
\begin{equation}\label{defnabla}
\left(\nabla_{Z}W\right)(g) := (L_{g})_{*}\left(Z\cdot\theta(W)
\right),
\end{equation}
where $Z\cdot\theta(W)$ denotes the derivative along the vector
$Z$ of the $\g$-valued function $\theta(W)$.
 Let us denote by $T$ and $R$
the torsion and the curvature of $\nabla$. By definition~:
$$
T(U, V) := \nabla_{U}V - \nabla_{V}U - [U, V]_{\mathfrak{X}}
$$
$$\!\!\!\textrm{and}\qquad R_{U, V}U := \nabla_{V}\nabla_{U}U - \nabla_{U}\nabla_{V}U -
\nabla_{[V, U]_{\mathfrak{X}}}.
$$
By \eqref{comUV}, one has
$$
\nabla_{U}V = \nabla_{V}U  + T(U, V),
$$
hence
$$ \nabla_{U}\left(\nabla_{U}V\right) = \nabla_{U}\nabla_{V}U
+ \nabla_{U}T(U, V).
$$
But the curvature tensor vanishes, hence   \eqref{comUV} and
\eqref{unablau} imply
$$
 \nabla_{U}\nabla_{V}U =  \nabla_{V}\nabla_{U}U -
\nabla_{[V, U]_{\mathfrak{X}}} = 0.
$$
Consequently one has
$$
\nabla_{U}\left(\nabla_{U}V\right) = \nabla_{U}T(U, V).
$$
By the expression \eqref{defnabla} of the connection, one has
$$
T(U, V)\left(\Phi(t, 0)\right) = \left( \nabla_{U}V -
\nabla_{V}U\right)(\Phi(t, 0)) = \left(L_{\Phi(t,
0)}\right)_{*}\left(U\cdot\theta(V) - V\cdot\theta(U)\right).
$$
Let us recall that the torsion is a tensor, hence $T(U,
V)\left(\Phi(t, 0)\right)$ does not depend on the extensions of
the vectors $U\left(\Phi(t, 0)\right)$ and $V\left(\Phi(t,
0)\right)$ into vector fields. Using the left-invariant extensions
of these two vectors one see easily that by the very definition of
the bracket in the Lie algebra $\g$ one has
$$
T(U, V)(\Phi(t, 0)) = -\left(L_{\Phi(t,
0)}\right)_{*}\left[\theta(U), \theta(V)\right].
$$
Whence
$$
\nabla_{U}\left(\nabla_{U}V\right) = -\nabla_{U}\left(L_{\Phi(t,
0)}\right)_{*}\left[\theta(U), \theta(V)\right] = \left(L_{\Phi(t,
0)}\right)_{*} \frac{d}{dt}\left[\theta(U), \theta(V)\right].
$$
 Now, along $\Phi(t, 0)$, the vector $\theta(U)$ is the
constant vector $X$, and $\theta(V) = v(t)$. It follows that
$$
\frac{d^{2}v(t)}{dt^2}  = \left(L_{\Phi(t,
0)}^{-1}\right)_{*}\nabla_{U}\left(\nabla_{U}V\right) = -
\left(L_{\Phi(t, 0)}^{-1}\right)_{*}\nabla_{U}\left(L_{\Phi(t,
0)}\right)_{*}\left[X , \theta(V)\right] = - \nabla_{U}[X,
\theta(V)].
$$
This leads to the following differential equation
$$
\frac{d^{2}v(t)}{dt^2}  = - \left[X, \frac{dv}{dt}\right]
$$
with initial conditions $ v(0) = 0 $ and $ \frac{dv}{dt}_{|t=0} =
Y.$ A first integration leads to $$\frac{dv}{dt} =
e^{-t\ad(X)}(Y)$$ and a second to $$v(t) = \left(\frac{1 -
e^{-t\ad(X)}}{\ad(X)}\right)(Y).$$ So the result follows from the
identity $v(1) =
\left(L_{\exp(X)}^{-1}\right)_{*}\left(d_{X}\exp\right)(Y)$.\cqfd

\begin{cor}\label{diffeo}
The exponential map is a diffeomorphism from
$\mathcal{S}_{2}(\Hc)$ onto $\mathcal{P}$.
\end{cor}

\proofc \ref{diffeo}:\\
For $X$ in $\mathcal{S}_{2}(\Hc)$, let us define the following
map~:
$$
\begin{array}{llll}
\tau_{X}~: &\mathcal{S}_{2}(\Hc) &\longrightarrow &
\mathcal{S}_{2}(\Hc)\\
&  Y & \mapsto &\tau_{X}(Y) := L_{\exp(-\frac{X}{2})}
R_{\exp(-\frac{X}{2})} d_{X}\exp(Y).
\end{array}
$$
Using the notation $D_{X} := \ad(X)$ and
$$
\frac{\exp \left(\frac{D_{X}}{2}\right) - \exp\left(
-\frac{D_{X}}{2}\right)}{D_{X}} =
\frac{\sinh\left(D_{X}/2\right)}{D_{X}/2} = \sum_{n = 0}^{+\infty}
\frac{(D_{X}/2)^{2n}}{(2n+1)!},
$$
we have as a direct consequence of formula \eqref{difexp} in
Proposition \ref{log} that for all $Y$ in $\mathcal{S}_{2}(\Hc)$
$$
\tau_{X}(Y) = \frac{\sinh(D_{X}/2)}{D_{X}/2}(Y),
$$
 Every $X$ in
$\mathcal{S}_{2}(\Hc)$ is a compact self-adjoint operator on
$\mathcal{H}$. Denote by $\{ \lambda_{i} \}_{i \in \N}$ the
spectrum of $X$, composed of real numbers such that $\sum_{i \in
\N} \lambda_{i}^{2} < + \infty$. The spectrum of $D_{X}$ acting on
$\mathfrak{gl}_{2}(\Hc)$ is then the set $\{ \lambda_{i} -
\lambda_{j}, i, j \in \N\}$, and the spectrum of $\tau_{X}$ is the
set:
$$
\left\{ \frac{\sinh(\frac{\lambda_{i}-
\lambda_{j}}{2})}{\frac{(\lambda_{i} -
    \lambda_{j})}{2}}, i, j \in \N\right\}.
$$
Since $$1 \leq \frac{\sinh(\frac{\lambda_{i}-
\lambda_{j}}{2})}{\frac{(\lambda_{i} -
    \lambda_{j})}{2}} \leq \frac{\sinh 2 \|X\|_{2}}{\|X\|_{2}},
$$
it follows that $\tau_{X}$ is one-one on $\mathcal{S}_{2}(\Hc)$
 and bounded.
Since the map that takes a formal series $f(\textrm{x})$ to the
operator $f(D_{X})$ of $B(\Hc)$ is a morphism of rings, the
inverse of $\tau_{X}$  is the operator given by:
$$
\begin{array}{llll}
\tau_{X}^{-1}:& \mathcal{S}_{2}(\Hc)  &\rightarrow &\mathcal{S}_{2}(\Hc)\\
              & Y & \mapsto & \frac{D_{X}/2}{\sinh D_{X}/2}(Y),
\end{array}
$$
whose norm is bounded by 1. Thus $\tau_{X}$ is an isomorphism of
$\mathcal{S}_{2}(\Hc)$ as well as $d_{X}\exp$. This implies that
$\exp$ is a local diffeomorphism on $\mathcal{S}_{2}(\Hc)$.
Moreover, since every $p$ in $\mathcal{P}$ admits an orthogonal
basis of eigenvectors with positive eigenvalues, the exponential
map from $\mathcal{S}_{2}(\Hc)$ to $\mathcal{P}$ is one-one and
onto, the inverse mapping being given by the logarithm. Therefore
$\exp$ is a diffeomorphism from $\mathcal{S}_{2}(\Hc)$ onto
$\mathcal{P}$. \cqfd

\begin{defe}{\rm
Let $G$ be a Banach Lie group. An homogeneous space $M = G/K$ is
called \emph{reductive} if the Lie algebra $\mathfrak{g}$ of $G$
splits into a direct sum $\mathfrak{g} = \mathfrak{k} \oplus
\mathfrak{m}$, where $\mathfrak{k}$ is the Lie algebra of $K$, and
$\mathfrak{m}$ an $\textrm{Ad}(K)$-invariant complement. A
reductive homogeneous space is called \emph{locally symmetric} if
the commutation relation $[\mathfrak{m}, \mathfrak{m}] \subset
\mathfrak{k}$ holds.}
\end{defe}

A locally symmetric space is a particular case of a
\emph{naturally reductive} space (see Definition 7.84 page~196 in
\cite{Bes}, Definition 23 page~312 in \cite{ON}, or Proposition
5.2 page~125 in \cite{Ar03} and the definition that follows). In
the finite-dimensional setting, the geodesics of a naturally
reductive space are orbits of one-parameter subgroups of $G$ (see
Proposition 25 page 313 in \cite{ON} for a proof of this fact).
The symmetric case is also treated in Theorem 3.3 page~173 in
\cite{Hel}.  Its infinite-dimensional version has been given in
Example 3.9 in \cite{Ne02c}. The proof we give below is based on
the notion of homogeneous connection.

\begin{prop}\label{genial}
Let $M = G/K$ be a locally symmetric homogeneous space under a
Banach Lie group $G$. Then, a geodesic of $M$ starting at $o = eK$
is given by
$$
\gamma(t) = \left(\exp t\a\right)\cdot o, \qquad \a \in
\mathfrak{m}.
$$
\end{prop}

\dem \ref{genial}:\\
Every element $\a$ in $\g$ generates a vector field $X^{\a}$ on
the homogeneous space $M = G/K$.
For every $x = g\cdot o$, $g \in G$, the Lie algebra $\g$ splits
into $\g = \mathfrak{k}_{x} \oplus \mathfrak{m}_{x}$, where
$\mathfrak{k}_{x} := \textrm{Ad}(g)(\mathfrak{k})$ is the Lie
algebra of the isotropy group at $x$ and where $\mathfrak{m}_{x}
:= \textrm{Ad}(g)(\mathfrak{m})$ can be identified with the
tangent space $T_{x}M$ of $M$ at $x$ by the application $\a
\mapsto X^{\a}(x)$. The homogeneous connection $\hat{\nabla}$ on
the tangent space of $M$ is defined as follows.
 For every element $\a$ in $\mathfrak{m}_{x}$ and every vector
field $X$ on $M$, one has
 \begin{equation}\label{connexhom}
\hat{\nabla}_{X^{\a}(x)}X = \left(\mathcal{L}_{X^{\a}}X\right)(x)
=  [X^{\a}, X]_{\mathfrak{X}}
 \end{equation}
where $\mathcal{L}$ denotes the Lie derivative and
$[\cdot\,,\,\cdot]_{\mathfrak{X}}$ the bracket of vector fields.
(The reader can check that \eqref{connexhom} defines indeed a
connection on the tangent bundle of $M$.) For $\a$ in
$\mathfrak{m}_{x}$ and $\b$ in $\g$, one has~:
$$
\hat{\nabla}_{X^{\a}(x)}X^{\b} =  [X^{\a},
X^{\b}]_{\mathfrak{X}}(x) = - X^{[\a, \b]}(x).
$$
The torsion of the connection $\hat{\nabla}$ is given by
$$
T^{\hat{\nabla}}(X^{\a}, X^{\b}) = \hat{\nabla}_{X^{\b}}X^{\a} -
\hat{\nabla}_{X^{\a}}X^{\b} - [X^{\a}, X^{\b}]_{\mathfrak{X}} = -
X^{[\a, \b]}.
$$
It follows that for a locally symmetric homogeneous space, the
homogeneous connection is torsion free since for $\a$ and $\b$ in
$T_{x}M = \mathfrak{m}_{x}$, $[\a, \b]$ belongs to the isotropy
$\mathfrak{k}_{x}$ thus $X^{[\a, \b]}$ vanishes. On the other
hand, it follows from definition \eqref{connexhom} that the
covariant derivation of any tensor field $\Phi$ along $Y \in
T_{x}M$ is the Lie derivative of $\Phi$ along the vector field
$X^{\a}$ where $\a \in \mathfrak{m}_{x}$ is such that $Y =
X^{\a}(x)$. Thus the homogeneous connection preserves every
$G$-invariant Riemannian metric. Consequently $\hat{\nabla}$ is
the Levi-Civita connection of every $G$-invariant Riemannian
metric on $M = G/H$. To see that for $\a \in \mathfrak{m}$, the
curve
$$ \gamma(t) = \left(\exp
t\a\right)\cdot o, \qquad \a \in \mathfrak{m}.
$$
is a geodesic, note that the equality $\a = \textrm{Ad}(\exp
t\a)(\a)$ implies that $\a$ belongs to the space
$\mathfrak{m}_{\gamma(t)}$ for all $t$. Hence from
$\dot{\gamma}(t) = X^{\a}(\gamma(t))$ it follows that
$\hat{\nabla}_{\dot{\gamma}(t)}\dot{\gamma}(t) =
\mathcal{L}_{X^{\a}}X^{\a}(\gamma(t)) = 0$. In other words
$\gamma$ is a geodesic of $M$. \cqfd

 \begin{rem} {\rm
 Note that for a weak
Riemannian metric on a Banach manifold, the existence of the
Levi-Civita connection is not guarantied in general (Koszul
formula defines a element in the dual of the tangent space, which
can not be represented by a vector in general). Hence the
symmetric homogeneous spaces form a class of Banach manifolds for
which the Levi-Civita connection exists. }
\end{rem}

\begin{lem}\label{formgeo}
The curve $\gamma(t) := \exp \left(t \log(p)\right)$, $( 0 \leq t
\leq 1)$ is the unique geodesic in
  $\mathcal{P}$ joining the identity $~o =\id~$ to the element $p
  \in \mathcal{P}$. More generally, the geodesic between any two
  points of $\mathcal{P}$ exists and is unique.
\end{lem}

\proofl \ref{formgeo}:\\
This follows from the same arguments as in \cite{Mostow}, or by
the general result stated in  Proposition \ref{genial}. Indeed the
commutation relation $[\mathcal{S}_2{\Hc}, \mathcal{S}_{2}(\Hc)]
\subset \mathfrak{u}_{2}(\Hc)$ implies that $\mathcal{P} =
\textrm{GL}_{2}(\Hc)/\textrm{U}_{2}(\Hc)$ is locally symmetric. It
follows that
$$\gamma(t) := \exp \left(t \log(p)\right) =
\left(\exp \left(\frac{t}{2} \log(p)\right)\right)\cdot o, \qquad
( 0 \leq t \leq 1),
$$
is a geodesic joining $~o = \id~$ to $p$. The uniqueness of this
geodesic follows from the injectivity of the exponential map,
since every other geodesic $\gamma_2$ joining $~o = \id~$ to $~p~$
is necessarily of the form $\gamma_{2}(t) = \exp
t\dot{\gamma_{2}}(0)$ by uniqueness of the geodesic starting at
$o$ with velocity $\dot{\gamma_{2}}(0)$. By the transitive action
of $\textrm{GL}_{2}(\Hc)$, there exists a unique geodesic
$\gamma_{p,q}$ joining two points $~p~$ and $~q~$, namely
$$\gamma_{p,q}(t) := p^{\frac{1}{2}}
\cdot \exp t \log\left(p^{-\frac{1}{2}} \cdot q\right) =
p^{\frac{1}{2}}\left( \exp t \log \left(p^{-\frac{1}{2}} q
p^{-\frac{1}{2}}\right)\right) p^{\frac{1}{2}}.$$ \cq

The following two Lemmas are standard results in the geometry of
non-positively curved spaces.

\begin{lem}\label{angle}
The Riemannian angle between two paths $f$ and $g$ intersecting at
$o$ is equal to the Euclidian angle between the two paths
$\log(f)$ and $\log(g)$ at $0$. Moreover, in any geodesic triangle
$ABC$ in $\mathcal{P}$,
$$
c^{2} \geq a^{2} + b^{2} - 2 a b \cos \widehat{ACB},
$$
where $a, b, c$ are the lengths of the sides opposite $A, B, C$
and $ \widehat{ACB}$ the angle at $A$.
\end{lem}

\demlem \ref{angle}:\\ This follows from the same arguments as in
\cite{Mostow}. The Al-Kashi inequality is also a direct
consequence of  Corollary 13.2 in \cite{Hel} page 73, since Lemma
\ref{formgeo} implies that $\mathcal{P}$ is a minimizing convex
normal ball.\cq


\begin{lem}\label{convex}
Let $\gamma_{1}(t)$ and $\gamma_{2}(t)$ be two constant speed
geodesics in $\mathcal{P}$. Then the distance in $\mathcal{P}$
between $\gamma_{1}(t)$ and $\gamma_{2}(t)$ is a convex function
of $~t$.
\end{lem} \cq


\newpage

\subsection{Geodesic subspaces of $\mathcal{P}$}

The following Theorem can be found in \cite{Mostow} in the
finite-dimensional case. It work as well in the
infinite-dimensional setting under an additional topological
hypothesis on $E$, which is that $E$ should be closed.

\begin{thm}[\cite{Mostow}]\label{equivgeospace}
Let $E$ be a \emph{closed} subspace of $\mathcal{S}_{2}(\Hc)$. The
following assertions are equivalent~:
\begin{enumerate}
\item \label{m1} $[ X, [X, Y]] \in E$ for all  $X, Y \in E$,
\item\label{m2} $e f e \in \exp E$ for all $e, f \in \exp E$,
\item \label{m3} $\exp E$ is a closed totally geodesic subspace of
  $\mathcal{P}$.
\end{enumerate}
\end{thm}

\begin{lem}\label{techu}:\\
Let $X$ be an element of $\mathcal{S}_{2}(\Hc)$. Define the
following maps
$$
a_{X}(A) =  A\!\cdot\!\exp X + \exp X\!\cdot\! A
$$
and $\gamma_{X} = (d_{X}\exp)^{-1} \circ a_{X}$. Then
$
\gamma_{X} = D_{X}\coth(D_{X}/2).
$
\end{lem}

\demlem \ref{techu}:\\
One has $a_{X} = e^{R_{X}} + e^{L_{X}},
$
where $R_{X}$ denotes right multiplication by $X$. By Proposition
\ref{log},
$$
\left(d_{X}\exp\right) = e^{L_{X}} \left(\frac{1 -
  e^{-\ad(X)}}{\ad(X)}\right) = e^{L_{X}} e^{-\frac{D_{X}}{2}} \frac{\sinh(\frac{D_{X}}{2})}{D_{X}/2}
= e^{L_{X}} e^{\frac{R_{X} -
L_{X}}{2}}\frac{\sinh(\frac{D_{X}}{2})}{D_{X}/2} =
e^{\frac{L_{X}}{2}}e^{\frac{R_{X}}{2}}\frac{\sinh(\frac{D_{X}}{2})}{D_{X}/2}.
  $$
Hence
$$
\left(d_{X}\exp\right)^{-1} = e^{-\frac{L_{X}}{2}}
e^{-\frac{R_{X}}{2}} \frac{ D_{X}/2 }{\sinh(\frac{D_{X}}{2})}.
$$
It follows that
$$
\gamma_{X} =  \frac{D_{X}}{\sinh(\frac{D_{X}}{2})}\,\,
e^{-\frac{L_{X}}{2}}\, e^{-\frac{R_{X}}{2}}\,\frac{e^{R_{X}} +
e^{L_{X}}}{2} = \frac{D_{X}}{\sinh(\frac{D_{X}}{2})}\,\,
\frac{e^{\frac{R_{X} - L_{X}}{2}} + e^{\frac{L_{X} -
R_{X}}{2}}}{2} = D_{X}\coth(D_{X}/2).
$$
\cq

\demthm \ref{equivgeospace}:

 $\ref{m1} \Rightarrow \ref{m2}:$ Suppose that $[X, [X, Y]] \in
E$ for all $X$, $Y$ in $E$. Let $f$ be an element in  $\exp E$,
$Y$ an element in  $E$ and $X$ the differentiable curve in
$S_{2}(\Hc)$ defined by
$$
X(t) = \log(\exp tY. f . \exp tY).
$$
Let us  show that $\exp X(t) = \exp tY. f . \exp tY$ belongs to
$\exp E$ for every $t$. One has
$$
{\frac{d}{dt}}_{|t = t_{0}} \exp X(t) = Y \exp X(t) + \exp X(t) Y,
$$
hence $X(t)$ satisfies the following differential equation~:
$$
\dot{X}(t) = (d_{X}\exp)^{-1}_{X(t)} a_{X(t)}(Y) =
\gamma_{X(t)}(Y) = D_{X(t)}\coth(D_{X}(t)/2)(Y).
$$
Note that  only even powers of $D$ are involved in the operator
$D\coth(D/2)$. Whence $X(t)$ belongs to the Banach space
 $E$ as soon as $X(0) \in E$. Moreover the flow of this vector field
 is
 defined for all  $t \in
\R$. Thus setting $t = 1$ and $Y = \log e$ with $e \in \exp E$,
give the result $e.f.e \in \exp E$.

 $\ref{m2} \Rightarrow \ref{m3}:$ Suppose that for all $e$ and
$f$ in $\exp E$, the product  $e.f.e$ belongs to $\exp E$. It
follows from Lemma \ref{formgeo} that $\exp E$ contains every
geodesic joining $\textrm{id}$ to an arbitrary element in $\exp
E$. Since the set of isometries of the form  $x \mapsto e. x. e$
with $e \in \exp E$ fixes $\exp E$ and acts transitively on it,
 every geodesic joining two points of $\exp E$ is contained in $\exp E$.
 In other words, the space  $\exp E$
is totally geodesic in $\mathcal{P}$.

 $\ref{m3} \Rightarrow \ref{m2}:$ Suppose that  $\exp E$ is a
closed totally geodesic subspace of $\mathcal{P}$. Let us consider
the symmetry $s_{p}$ with respect to $p \in \mathcal{P}$ defined
from $\mathcal{P}$ to $\mathcal{P}$ by $s_{p}: x \mapsto p x^{-1}
p$ with $p \in \mathcal{P}$.  Every geodesic of the form $t
\mapsto p^{-\frac{1}{2}}\exp(tX) p^{-\frac{1}{2}}$ is mapped to $t
\mapsto p^{-\frac{1}{2}}\exp(-tX) p^{-\frac{1}{2}}$ by $s_{p}$. It
follows that every geodesic containing   $p$ is stable under
$s_{p}$. Consequently if $\exp E$ is a totally geodesic subspace
of $\mathcal{P}$, then $s_{p}(\exp E) \subset \exp E$ for every $p
\in \exp E$. Let $\tau_{p}$ denote the isometry of $\mathcal{P}$
defined by $\tau_{p}(x) = p^{\frac{1}{2}} x p^{\frac{1}{2}}$. Then
$$
s_{p}.s_{p^{\frac{1}{2}}}(x) = p.(p^{\frac{1}{2}} x^{-1}
  p^{\frac{1}{2}})^{-1}.p = p^{\frac{1}{2}} x p^{\frac{1}{2}} =
  \tau_{p}(x).
$$
Whence for every $e$, $f$ in $\exp E$, $e.f.e = \tau_{e}(f) =
\sigma_{e}(\sigma_{e^{\frac{1}{2}}}(f)) \in \exp E$.

 $\ref{m2} \Rightarrow \ref{m1}$: Suppose that $e.f.e \in \exp E$
as soon as $e$, $f$ belong to $\exp E$. For $f$ in $\exp E$ and
$Y$ in $E$, let  $X$ be the differentiable curve in
$\mathcal{S}_{2}(\Hc)$ defined by
$$
X(t) = \log(\exp tY. f . \exp tY).
$$
Then $X(t)$ belongs to $E$ for all $t \in \R$, as well as its
derivative $\dot{X}(t)$. Therefore
$$
Z = \lim_{t \mapsto 0} \frac{\dot{X}(t) - \dot{X}(0)}{t^{2}}
$$
belongs to $E$ also. Since $\dot{X}(t) =
D_{X(t)}\coth(D_{X(t)}/2)(Y)$, one has
$$
Z = \lim_{t \mapsto 0}\left[\frac{(1 + (1/12)t^{2}D_{X(t)}^{2})Y -
Y}{t^{2}} +
  tW\right] = (1/12)t^{2}D_{X(0)}^{2}Y,
$$
where $W$ depends continuously on $t$. Hence $D_{X(0)}Y$ belongs
to $E$. Tacking $f = \exp X$ gives the result. \cqfdt

\subsection{Orthogonal projection on a geodesic
subspace}\label{sub3}

In the following, $E$ is a closed linear subspace of
$\mathcal{S}_{2}(\Hc)$ such that $[ X, [X, Y]] \in E$, for all $X,
Y \in E$. From corollary \ref{diffeo}, it follows that $\exp E$ is
closed in $\mathcal{P}$.

The proof of Mostow's decomposition theorem given in \cite{Mostow}
is based on the existence of an orthogonal projection from
$\mathcal{P}$ onto $\exp E$ which follows from compactness
arguments that can't be used in the infinite dimensional setting.
Here we use the completeness of $\exp E$ to obtain an analogous
result.

\begin{thm}\label{projection}
There exist a continuous orthogonal projection  from $\mathcal{P}$
onto $\exp E$, i.e. a continuous map $\pi$  satisfying
$\textrm{dist}(p, \exp E) = \textrm{dist}(p, \pi(p))$ and such
that the geodesic joining $p$ to $\pi(p)$ is orthogonal to every
geodesic starting from $\pi(p)$ and included in $\exp E$.
\end{thm}

\prooft \ref{projection}:\\
 Let $p$ be a element of $\mathcal{P}$. Denote by $\delta$ the
distance between $p$ and $\exp E$ in $\mathcal{P}$ and let
$\{e_{n} \}_{n \in \N}$ be a sequence in $\exp E$ thus that
$$
\textrm{dist}(p, e_{n})^{2} \leq \delta^{2} + \frac{1}{n}.
$$
Let us show that $\{ e_{n} \}_{n \in \N}$ is a Cauchy sequence in
$\exp E$. For this purpose, consider for $k > n$ the geodesic
$\gamma(t)$ joining $e_{n} =: \gamma(0)$ to $e_{k}:= \gamma(1)$.
This geodesic  lies in $\exp E$ since $\exp E$ is a geodesic
subspace of $\mathcal{P}$, and is of the form:
$$
\gamma(t) = e_{1}^{\frac{1}{2}} \exp(t H) e_{1}^{\frac{1}{2}},
$$
where $H$ belongs  to $E$. Denote by $e_{n, k}$ the middle of the
geodesic joining $e_{n}$ to $e_{k}$, i.e. $e_{n, k} =
e_{1}^{\frac{1}{2}} \exp (\frac{1}{2} H) e_{1}^{\frac{1}{2}}$. By
lemma \ref{angle} applied to the geodesic triangle joining $p$,
$e_{n}$ and $e_{n, k}$, we have:
$$
\textrm{dist}(p, e_{n})^{2} \geq \textrm{dist}(e_{n}, e_{n,
k})^{2} + \textrm{dist}(e_{n, k}, p)^{2} - 2 \textrm{dist}(e_{n},
e_{n, k}) \textrm{dist}(e_{n, k}, p) \cos \widehat{e_{n} e_{n, k}
p}.
$$
On the other hand, lemma \ref{angle} applied to the geodesic
triangle joining $p$, $e_{k}$ and $e_{n, k}$ gives:
$$
\textrm{dist}(p, e_{k})^{2} \geq \textrm{dist}(e_{k}, e_{n,
k})^{2} + \textrm{dist}(e_{n, k}, p)^{2} - 2 \textrm{dist}(e_{k},
e_{n, k}) \textrm{dist}(e_{n, k}, p) \cos \widehat{e_{k} e_{n, k}
p}.
$$
By definition of $e_{n, k}$ we have: $\textrm{dist}(e_{k}, e_{n,
k}) = \textrm{dist}(e_{n}, e_{n,
  k})$. Moreover since the geodesic $\gamma$ is a smooth curve:
$$
\widehat{e_{k} e_{n, k} p} + \widehat{e_{n} e_{n, k} p} =
180^{\circ},
$$
and $\cos \widehat{e_{k} e_{n, k} p} = - \cos \widehat{e_{n} e_{n,
k}
  p}.$
Summing both inequalities, we obtain:
$$
\textrm{dist}(p, e_{n})^{2} + \textrm{dist}(p, e_{k})^{2} \geq 2
\textrm{dist}(e_{k}, e_{n, k})^{2} + 2 \textrm{dist}(e_{n, k},
p)^{2}.
$$
It follows that:
$$
\begin{array}{ll}
\textrm{dist}(e_{k}, e_{n, k})^{2} & \leq
\frac{1}{2}(\textrm{dist}(p, e_{n})^{2} + \textrm{dist}(p,
e_{k})^{2}) - \textrm{dist}(e_{n, k},
p)^{2}\\
& \leq \frac{1}{2}(\delta^{2} + \frac{1}{n} + \delta^{2} +
\frac{1}{k}) -
\delta^{2}\\
& \leq \frac{1}{2}(\frac{1}{n} + \frac{1}{k}).
\end{array}
$$
This yields that $\textrm{dist}(e_{n}, e_{k}) \leq
\sqrt{2}(\frac{1}{n} + \frac{1}{k})^{\frac{1}{2}}$ and
$\{e_{n}\}_{n
  \in \N}$ is a Cauchy sequence in $\exp E$.
Since $\exp E$ is closed in the complete space
$\mathcal{S}_{2}(\Hc)$, the sequence $\{ e_{n} \}_{n \in \N} $
converges to a element $\pi(p)$ in $\exp E$ satisfying:
$$
 \textrm{dist}(p, \pi(p)) = \textrm{dist}(p, \exp E).
$$

 Denote by $\alpha(t)$ the constant speed geodesic
 which satisfies $\alpha(0) = \pi(p)$ and $\alpha(1) = p$. By
uniqueness of the geodesic joining two points it follows that the
length of $\alpha $ is $\textrm{dist}(p, \exp E)$. The map $x
\mapsto (\pi(p))^{-\frac{1}{2}} x (\pi(p))^{-\frac{1}{2}}$ being
an isometry, the curve $(\pi(p))^{-\frac{1}{2}} \alpha
(\pi(p))^{-\frac{1}{2}}$ is a geodesic whose length is the
distance between $(\pi(p))^{-\frac{1}{2}} p
(\pi(p))^{-\frac{1}{2}}$ and $\exp E$, thus the projection of
$(\pi(p))^{-\frac{1}{2}} p (\pi(p))^{-\frac{1}{2}}$ onto $\exp E$
is $o$. From lemma \ref{formgeo} it follows that:
$$
(\pi(p))^{-\frac{1}{2}} \alpha(t) (\pi(p))^{-\frac{1}{2}} = \exp t
V,
$$
for some $V$ in $\mathcal{S}_{2}(\Hc)$. Since the length of $t
\mapsto \exp tV$ is $\|V\|$, $V$  is in $F$ and
$(\pi(p))^{-\frac{1}{2}} p (\pi(p))^{-\frac{1}{2}}$ is in $\exp
F$. Since $E \perp F$, by lemma \ref{angle},
$(\pi(p))^{-\frac{1}{2}} \alpha (\pi(p))^{-\frac{1}{2}}$ is
orthogonal at the identity to every curve starting at the identity
and contained in $\exp E$. Therefore $\alpha$ is orthogonal at
$\pi(p)$ to every curve starting at $\pi(p)$ and contained in
$\exp E$.

 To show that $\pi$ is continuous, denote by $\gamma(t)$
 (resp. $\alpha(t)$) the geodesic
joining a  points $p_{1}$ (resp. $p_{2}$) in $\mathcal{P}$ to its
projection on $\exp E$, with $\gamma(0) = \pi(p_{1})$ (resp.
$\alpha(0) = \pi(p_{2})$ ) and $\gamma(1) = p_{1}$ (resp.
$\alpha(1) = p_{2}$). By the negative curvature property stated in
Lemma \ref{convex}, the map $t \mapsto \textrm{dist}(\gamma(t),
\alpha(t))$ is convex. Since, for $t = 0$, $\gamma(t)$ and
$\alpha(t)$ are orthogonal to the geodesic joining $\pi(p_{1})$
and $\pi(p_{2})$, the minimum of the distance between $\gamma(t)$
and $\alpha(t)$ is reached for $t = 0$, and $\textrm{dist}(p_{1},
p_{2}) \geq \textrm{dist}(\pi(p_{1}), \pi(p_{2}))$. \cqfdt

\subsection{Proof of Mostow's Decomposition Theorem}

\begin{thm}\label{mom}
Let $E$ be a closed linear subspace of $\mathcal{S}_{2}(\Hc)$ such
that:
$$
[ X, [X, Y]] \in E, \qquad for~all\quad X, Y \in E,
$$
and let $F$ be its orthogonal in $\mathcal{S}_{2}(\Hc)$:
$$
F := E^{\perp} = \{ ~X \in \mathcal{S}_{2}(\Hc)~|~ \Tr XY = 0,~~
\forall Y \in E ~\}.
$$
For all self-adjoint positive-definite operator $A$ in
$\mathcal{S}_{2}(\Hc)$, there exist a unique element $e \in \exp
E$ and a unique element $f \in \exp F$ such that $A = e f e.$
Moreover the map defined from $\mathcal{P}$ to $\exp E \times \exp
F$ taking $A$ to $(e, f)$ is a homeomorphism.
\end{thm}

\prooft \ref{mom}:\\
Denote by $\Upsilon$ the map from $\exp E \times \exp F$ to
$\mathcal{P}$ that takes $(e, f)$ to $e f e$.

 Let us show that $\Upsilon$ is one-one. Suppose that $(e_{1},
f_{1})$ and $(e_{2}, f_{2})$ are elements of $\exp E \times \exp
F$ such that $e_{1} f_{1} e_{1} = e_{2} f_{2} e_{2}$. Consider the
geodesic triangle joining  $e_{1} f_{1} e_{1}$, $e_{1}^{2}$ and
$e_{2}^{2}$. By Theorem \ref{equivgeospace}, $\exp E$ is a
geodesic subspace of $\mathcal{P}$. Thus the geodesic joining
$e_{1}^{2}$ to $e_{2}^{2}$ lies in $\exp E$. On the other hand the
geodesic joining $e_{1} f_{1} e_{1}$ to $e_{1}^{2}$ lies in $e_{1}
\exp F \,e_{1}$. Since $E$ is perpendicular to $F$ at zero, $\exp
E$ is perpendicular to $\exp F$ at the identity by lemma
\ref{angle}. Now the map taking $x$ to $e_{1} x\, e_{1}$ is an
isometry, thus $e_{1} \exp F\, e_{1}$ is perpendicular to $e_{1}
\exp E\, e_{1} = \exp E$ at $e_{1}^{2}$. Hence the angle at
$e_{1}^{2}$ of the above geodesic triangle is $90^{\circ}$.
Similarly, the angle at $e_{2}^{2}$ is $90^{\circ}$ since it is
formed by the geodesic joining $e_{2}^{2}$ to $e_{2} f_{2} e_{2} =
e_{1} f_{1} e_{1}$ which lies in $e_{2} \exp F\, e_{2}$ and the
geodesic joining $e_{2}^{2}$ to $e_{1}^{2}$ which lies in $\exp
E$. Denoting by $a$ the length of the side of the geodesic
triangle joining $e_{1}^{2}$ to $e_{2}^{2}$, $b$ the length of the
side joining $e_{1} f_{1} e_{1} $ to $e_{1}^{2}$ and $c$ the
length of the side joining $e_{1} f_{1} e_{1}$ to $e_{2}^{2}$,
ones has $c^{2} \geq b^{2} + a^{2}$ and $b^{2} \geq c^{2} + a^{2}$
by lemma \ref{angle}. This implies that $a = 0$ and $e_{1}^{2} =
e_{2}^{2}$. It follows that $e_{1} = e_{2}$ and $f_{1} = f_{2}$.

 Let us show that $\Upsilon$ is onto. Consider $p$ in
$\mathcal{P}$. By Theorem \ref{projection}, the geodesic joining
$p$ to $\pi(p) \in \exp E$ is orthogonal to every geodesic
starting at $\pi(p)$ and contained in $\exp E$. Denote by $\gamma$
the geodesic satisfying $\gamma(0) = o$ and $\gamma(1) =
(\pi(p))^{-\frac{1}{2}} p (\pi(p))^{-\frac{1}{2}}$.  Since $x
\mapsto (\pi(p))^{-\frac{1}{2}} x (\pi(p))^{-\frac{1}{2}}$ is an
isometry,  $\gamma$  is orthogonal to every geodesic starting at
the identity and contained in $(\pi(p))^{-\frac{1}{2}} \exp E
(\pi(p))^{-\frac{1}{2}} = \exp E$. By lemma \ref{angle}, $\gamma$
is tangent to $F = E^{\perp}$ at the identity and since it is of
the form $t \mapsto \exp t H$ by lemma \ref{formgeo}, we have $H$
in $F$. It follows that $\gamma(1) = \exp H =
(\pi(p))^{-\frac{1}{2}} p (\pi(p))^{-\frac{1}{2}}$ is in $\exp F$.
Therefore $p = e f e$ with $e := (\pi(p))^{-\frac{1}{2}}$ in $\exp
E$ and $f := (\pi(p))^{-\frac{1}{2}} p (\pi(p))^{-\frac{1}{2}}$ in
$\exp F$ and $\Upsilon$ is onto.

 The continuity of the map that takes $p$ to $(e, f) \in \exp E
\times \exp F$ with $p = e f e$ follows directly from the
continuity of the projection $\pi$. \cqfdt

\begin{thm}[Mostow's Decomposition]\label{glm}
Let $E$ and $F$ be as in Theorem \ref{mom}. Then
$\textrm{GL}_{2}(\Hc)$ is homeomorphic to the product
$U_{2}(\Hc)\cdot\exp F\cdot\exp E$.
\end{thm}

\prooft \ref{glm}:\\
Denote by $\Theta$ the map from $U_{2}(\Hc) \times \exp E \times
\exp F$ to $\textrm{GL}_{2}(\Hc)$ that takes $(k, f, e)$ to $k f
e$.

 Let us show that $\Theta $ is one-one. Suppose that $a = k_{1} f_{1}
e_{1} = k_{2} f_{2} e_{2}$ with $(k_{1}, f_{1}, e_{1})$ and
$(k_{2}, f_{2}, e_{2})$ in $U_{2}(\Hc) \times \exp E \times \exp
F$. We have
$$
a^{*}a = e_{1} f^{2}_{1} e_{1} = e_{2} f_{2}^{2} e_{2}.
$$
Since $f_{1}^{2}$ and $f_{2}^{2}$ are in $\exp F$, by Theorem
\ref{mom}, it follows that $e_{1} = e_{2}$ and $f_{1}^{2} =
f_{2}^{2}$. Thus $f_{1} = f_{2}$ and $k_{1} = k_{2}$.

 Let us show that $\Theta$ is onto. Consider $x$ in $\textrm{GL}_{2}(\Hc)$.
$x^{*}x$ is an element of $\mathcal{P}$ and by Theorem \ref{mom},
there exist $e \in \exp E $ and $f \in \exp f$ such that $x^{*}x =
e f^{2} e$. Let $k$ be $x (f e)^{-1}$. We have:
$$
k^{*}k = (f e)^{-1*} x^{*}x (fe)^{-1} = f^{-1} e^{-1} e f^{2} e
e^{-1} f^{-1} = \textrm{ id }.
$$
Thus $k$ is in $U_{2}(\Hc)$ and $x = k f e$.

 The continuity of the map that takes $x$ in $\textrm{GL}_{2}(\Hc)$ to $(k,
f, e) $ in $U_{2}(\Hc) \times \exp F \times \exp E$ follows from
the continuity of the map that takes $x$ to $x^{*}x$ and from
theorem \ref{mom}. \cqfdt

\begin{thm}[= Theorem \ref{debut}]\label{mostogro}
 Let $G$ be a semi-simple connected
$L^{*}$-group of compact type with Lie algebra  $\mathfrak{g}$,
$G^{\C}$ the connected $L^{*}$-group with Lie algebra
$\mathfrak{g}^{\C} := \mathfrak{g} \oplus \ie \g$, $E$ a closed
subspace of  $~\ie\g$ such that
$$
\left[ X, [X, Y]\right] \in E, \qquad for ~all ~~X, Y \in E,
$$
and $F$ the orthogonal of $E$ in $\ie\g$. Then $G^{\C}$ is
homeomorphic to the product  $G \cdot \exp F \cdot \exp E$.
\end{thm}

\demthm \ref{mostogro}:\\
Since $\g^{\C}$ is a semi-simple $L^{*}$-algebra, it decomposes
into a Hilbert sum of closed  $*$-stable simple ideals $\g_{j}$,
for $j$ in some set $J$ (see \cite{Schu1}).
Since every simple $L^{*}$-algebra $\g_{j}$ is a subalgebra of the
$L^*$-algebra of Hilbert-Schmidt operators on some Hilbert space
$\mathcal{H}_{j}$, $\g^{\C}$ is a $L^{*}$-subalgebra of
$\mathfrak{gl}_{2}(\Hc)$ where $\Hc$ is the Hilbert sum of
$\mathcal{H}_{j}$, $j \in J$. The complex $L^*$-group $G^{\C}$ is
therefore an $L^{*}$-subgroup of $\textrm{GL}_{2}(\Hc)$. Since
$G^{\C}$ is connected, $G^{\C}$ is generated by a neighborhood of
the unit element. The exponential mapping is a local
diffeomorphism from a neighborhood of $0$ in $\g^{\C}$ onto a
neighborhood of the unit in  $G^{\C}$. Since $\g^{\C}$ is
$*$-stable and since $(\exp X)^{*} = \exp X^{*}$, $G^{\C}$ is also
$*$-stable. Let $x \in G^{\C}$. Mostow's Decomposition Theorem
implies that $x$ can be written as $x = k.f.e$ with  $k \in
\textrm{U}_{2}(\Hc)$, $e \in \exp E$ and $f \in \exp F'$, where
$F'$ is the orthogonal of $E$ in $\mathcal{S}_{2}(\Hc)$. Since
$G^{\C}$ is $*$-stable, it contains $x^{*}x = e f^{2} e$. But
$x^*x$ is an Hermitian positive-definite operator in $G^{\C}$,
thus belongs to $\exp \mathcal{S}_{2}(\Hc) \cap G^{\C} = \exp
\ie\g$. Since $f^{2} = e^{-1} x^{*}x e^{-1}$, $f^{2}$ belongs also
to $\exp \ie\g$ hence to $\exp F$, as well as its square root $f$.
It follows that $k = x e^{-1} f^{-1}$ belongs to $G^{\C} \cap
\textrm{U}_{2}(\Hc) = G$. \cqfdt

\begin{cor}\label{lkj}
Let $G$ be a connected semi-simple  $L^{*}$-group of compact type
with Lie algebra $\mathfrak{g}$ and $G^{\C}$ the connected
$L^{*}$-group with  Lie algebra $\mathfrak{g}^{\C}= \g \oplus \ie
\g$. Given an $L^*$-subalgebra  $\k$ of $\g$, denote by
$\mathfrak{m}$ the orthogonal of $\k$ in $\g$. Then $G^{\C}$ is
homeomorphic to   $G \cdot \exp \ie\mathfrak{m} \cdot \exp \ie\k$.
\end{cor}

\demcor \ref{lkj}:\\
This is a direct consequence of Theorem \ref{mostogro} since
$[\ie\k, [\ie\k, \ie\k ]] \subset \ie\k$. \cqfd

\newpage
\section{Complexification and cotangent bundle
 of  coadjoint orbits}\label{section2}

\subsection{Finite-dimensional Theorem}

 Mostow's Decomposition Theorem implies the following.
\begin{thm}\label{projectionorbite}
Let $G$ be a semi-simple compact Lie-group with Lie algebra
$\mathfrak{g}$ and denote by $G^{\C}$ the unique complex group
with Lie algebra $\g:= \g \oplus \ie \g$ such that $G$ injects
into $G^{\C}$ and such that this injection induces the natural
injection $\g \hookrightarrow \g^{\C}$. Let $x$ be an element in
$\g$. Denote by $\mathcal{O}_{x}$ the adjoint orbit of $x$ under
$G$, by $\mathcal{O}_{x}^{\C}$ the adjoint orbit of $x$ under
$G^{\C}$, and by $p~: T\mathcal{O}_{x} \rightarrow
\mathcal{O}_{x}$ the tangent bundle of the compact orbit
$\mathcal{O}_{x}$. Then there exists a $G$-equivariant projection
$ \pi~: \mathcal{O}_{x}^{\C} \twoheadrightarrow \mathcal{O}_{x} $
and a $G$-equivariant homeomorphism  $ \Phi$ from the tangent
space $T\mathcal{O}_{x}$ onto the complex orbit $
\mathcal{O}_{x}^{\C} $ which commutes with the projections $p$ and
$\pi$.
\end{thm}

\begin{lem}\label{nonrecoupement}
Let $~z~$ be an element of the compact adjoint orbit
$\mathcal{O}_{x}$. Denote by $~\mathfrak{k}_{z}~$ the Lie algebra
of the stabilizer of $~z~$, and by $~\mathfrak{m}_{z}~$ the
orthogonal of $~\mathfrak{k}_{z}~$ with respect to the Killing
form of $~\g~$. Then
$$
\exp\left(\ie \mathfrak{m}_{z}\right)\cdot z ~\cap ~\g ~= ~\{z\}
$$
\end{lem}

\demlem \ref{nonrecoupement}:\\
Let $\a \in \mathfrak{m}_{z}$ be such that $\exp\left(\ie
\a\right)\cdot z$ belongs to $\g$. Recall that
$$
\exp\left(\ie \a\right)\cdot z = \textrm{Ad}\left(\exp\ie
\a\right)(z) = e^{\ie\ad\a}(z) = \sum_{n=1}^{+\infty}
\frac{(\ie~\ad\a)^{n}}{n!}(z) = \cosh\left(\ie~\ad(a)\right)(z) +
\sinh\left(\ie~\ad(a)\right)(z),
$$
where $\cosh\left(\ie~\ad(a)\right)(z)$ belongs to $\g$ and
$\sinh\left(\ie~\ad(a)\right)(z)$ to $\ie \g$.  Hence the
condition $\exp\left(\ie \a\right)\cdot z \in \g$ reads
$$
0 =  \sinh(\ie~\ad\a)(z).
$$
Since the operator $\ad(\a)$ is skew-symmetric with respect to the
Killing form, the $\C$-extension $\g^{\C}$ of $\g$ splits into a
sum of eigenspaces $\g_{\lambda_{j}}$ of $\ad(\a)$ with
eigenvalues $\ie\lambda_{j}$, where $\lambda_{j}\in \mathbb{\R}$.
Let $z = \sum_{j \in J} z_{\lambda_{j}}$ be the decomposition of
$z$ with respect to the direct sum $\g^{\C} = \sum_{j\in
J}\g_{\lambda_{j}}$. One has
$$\ie~\ad(\a)(z) = -\sum_{j\in J}
\lambda_{j}~z_{\lambda_{j}}$$ and
$$
\sinh(\ie~\ad\a)(z) = - \sum_{j\in
J}\sinh(\lambda_{j})~z_{\lambda_{j}}.
$$
It follows that $\sinh(\ie~\ad\a)(z)$ vanishes if and only if
$\sinh(\lambda_{j})~z_{\lambda_{j}}$ vanishes for all $j\in J$, or
equivalently if and only if for $\lambda_{j} \neq 0$,
$z_{\lambda_{j}} = 0$. Thus $z$ belongs to the eigenspace $\g_{0}$
which is the kernel of $\ad(\a)$. But the equation $[\a, z] = 0$
implies that $\a = 0$ since $\a$ belongs to $\mathfrak{m}_{z}$ by
hypothesis. Consequently $ \exp\left(\ie
\mathfrak{m}_{z}\right)\cdot z ~\cap ~\g $ reduces to $z$ and the
Lemma is proved.
 \cq
\\
\\

\demthm \ref{projectionorbite}:\\
Let us first show that every $y$ in the complex adjoint orbit
$\mathcal{O}_{x}^{\C}$ can be written uniquely as
\begin{equation}\label{ecriture}
y = \exp (\ie \a)\cdot z, \end{equation} where $z$ belongs to
$\mathcal{O}_{x}$ and $\a$ to $\mathfrak{m}_{z}$. Since $y$
belongs to the complex orbit of $x$, there exists $g\in G^{\C}$
such that $y = g\cdot x = \textrm{Ad}(g)(x)$. By Mostow's
Decomposition Theorem, there exist $u \in G$,
$\mathfrak{b}\in\mathfrak{m}_{x}$ and $\mathfrak{c} \in
\mathfrak{k}_{x}$ such that $g = u \exp \ie \mathfrak{b} \exp \ie
\mathfrak{c}$. It follows that $y = (u \exp \ie \b)\cdot x$ since
$\exp \ie \mathfrak{k}_{x}$ acts trivially on $x$. But $u \exp \ie
\b = \exp\left(\ie \textrm{Ad}(u)(\b)\right) u$. Hence $y = \exp
(\ie \a)\cdot z$ with $z := u\cdot x \in \mathcal{O}_{x}$ and $\a
:= \textrm{Ad}(u)(\b) \in \mathfrak{m}_{z}$. This proves the
existence of the expression \eqref{ecriture}. In order to proof
uniqueness, let us suppose that
$$
\exp\left(\ie \a_{1}\right)\cdot z_{1} = \exp\left(\ie
\a_{2}\right)\cdot z_{2},
$$
for some $z_{1}$, $z_{2}$ in $\mathcal{O}_{x}$, some $\a_{1}$ in
$\mathfrak{m}_{z_{1}}$ and some $\a_{2}$ in
$\mathfrak{m}_{z_{2}}$. One has~:
\begin{equation}\label{bla}
\exp\left(-\ie \a_{2}\right)\exp\left(\ie \a_{1}\right)\cdot z_{1}
= z_{2}.
\end{equation}
By Mostow's Decomposition Theorem, there exists $u'$ in $G$,
$\a_{3}$ in $\textrm{m}_{z_{1}}$ and $\b_{3}$ in
$\mathfrak{k}_{z_{1}}$ such that
\begin{equation}\label{pdt}
\exp\left(-\ie \a_{2}\right)\exp\left(\ie \a_{1}\right) =
u'\exp(\ie \a_{3})\exp(\ie \b_{3}).
\end{equation}
Thus equation \eqref{bla} becomes
$$
u' \exp(\ie \a_{3})\cdot z_{1} = z_{2},
$$
or equivalently
$$
 \exp(\ie \a_{3})\cdot z_{1} = (u')^{-1}\cdot z_{2},
$$
But $(u')^{-1}\cdot z_{2}$ belongs to $\mathcal{O}_{x}$ since
$z_{2}$ is an element of $\mathcal{O}_{x}$ and $u' \in G$. Whence
Lemma \ref{nonrecoupement} implies that
$$
\exp(\ie \a_{3})\cdot z_{1} = (u')^{-1}\cdot z_{2} = z_{1}.
$$
It follows that $\a_{3}$ stabilizes $z_{1}$, hence vanishes
because it belongs to  $\mathfrak{k}_{z_{1}} \cap
\mathfrak{m}_{z_{1}} = \{0\}$, and that $z_{2} = u'\cdot z_{1}$.
Therefore equation \eqref{pdt} becomes
$$
\exp\left(-\ie \a_{2}\right)\exp\left(\ie \a_{1}\right) =
u'\exp(\ie \b_{3}).
$$
It follows that
$$
\exp\left(\ie \a_{1}\right) = \exp\left(\ie
\a_{2}\right)u'\exp(\ie \b_{3}) = u'~\exp\left(\ie
\textrm{Ad}\left((u')^{-1}\right)(\a_{2})\right)~\exp(\ie \b_{3}),
$$
where  $\textrm{Ad}\left((u')^{-1}\right)(\a_{2})$ belongs to
$\textrm{Ad}\left((u')^{-1}\right)\left(\mathfrak{m}_{z_{2}}\right)
= \mathfrak{m}_{z_{1}}$. By uniqueness of Mostow's decomposition,
one  has
$$
u' = e, \qquad \a_{2} = \a_{1}, \qquad \textrm{and}~~\b_{3} = 0.
$$
Thus uniqueness of \eqref{ecriture} is proved. The projection
$\pi$ is defined as follows~:
$$
\begin{array}{lcll}
\pi~:&   \mathcal{O}_{x}^{\C} & \twoheadrightarrow &
\mathcal{O}_{x}\\
& y = \exp (\ie \a)\cdot z   & \mapsto & z,
\end{array}
$$
where in the expression of $y$, the element $z$ is supposed to
belong to $\mathcal{O}_{x}$ and $ \a$ to $\mathfrak{m}_{z}$. The
$G$-equivariance of the projection $\pi$ is a direct consequence
of the identity
$$
u\cdot y = \exp \left(\ie \textrm{Ad}(u)(\a)\right)\cdot (u \cdot
z),
$$
and of the relation $\textrm{Ad}(u)(\mathfrak{m}_{z}) =
\mathfrak{m}_{u\cdot z}$. Let us recall that the tangent space to
$\mathcal{O}_{x}$ at $z$ can be identified with
$\mathfrak{m}_{z}$. Define $\Phi$ by
$$
\begin{array}{llll}
\Phi~:& T\mathcal{O}_{x} & \rightarrow & \mathcal{O}_{x}^{\C}\\
  &    \a \in \mathfrak{m}_{z} = T_{z}\mathcal{O}_{x} & \mapsto &
  \exp(\ie \a)\cdot z.
\end{array}
$$
The $G$-equivariance of $\Phi$ is clear. It is an homeomorphism of
fiber bundles since Mostow's Decomposition Theorem is an
homeomorphism.
 \cqfdt

\subsection{Infinite-dimensional Theorem}

An infinite-dimensional analogue of Theorem \ref{projectionorbite}
is given in Theorem \ref{infiniteanal}. In order to state it, let
us introduce some notation. In the following, $G$ will denote a
\emph{simple} $L^{*}$-group of compact type, that is an Hilbert
Lie group  such that
\begin{equation}\label{*}
\langle [x\,,\, y]\,,\,z\rangle = -\langle y\,,\,[x\,,\,z]\rangle.
\end{equation}
It follows from the classification Theorem of simple $L^*$-groups
of compact type given in \cite{Bal}, \cite{Har2}, or \cite{Uns2},
that $G$ is a group of Hilbert-Schmidt operators on a certain
Hilbert space $\Hc$. Denote by $\g$ the Lie algebra of $G$ and by
$G^{\C}$ the complex $L^{*}$-group with Lie algebra $\g:= \g
\oplus i \g$ characterized by the property that $G$ injects into
$G^{\C}$ and that this injection induces the natural injection $\g
\hookrightarrow \g^{\C}$. According to Theorem 4.4 in \cite{Nee2},
every derivation of $\g$ \emph{with closed range} is
diagonalizable on $\g^{\C}$ and represented by the bracket with
some skew-Hermitian operator $D$ with finite spectrum. Denote by
$\mathcal{O}^{\C}$ the orbit of $0$ under the affine adjoint
action of $G^{\C}$ defined by
$$
\begin{array}{llll}
\textrm{Ad}_{D}~:& G^{\C} \times \g^{\C} & \longrightarrow & \g^{\C}\\
& (g\,,\, y) & \longmapsto &  g~y~g^{-1} + g~D~g^{-1} - D,
\end{array}
$$
and by $\mathcal{O}$ the affine adjoint orbit of $0$ under the
restriction of $\textrm{Ad}_{D}$ to $G$.

\begin{thm}\label{infiniteanal}
There exists a $G$-equivariant projection $\pi$ from
$\mathcal{O}^{\C}$ onto $\mathcal{O}$ and a $G$-equivariant
homeomorphism $\Phi$ from the tangent space to the orbit
$\mathcal{O}$ of compact type onto the complex orbit
$\mathcal{O}_{x}^{\C}$ which commutes with the projection $p$ and
$\pi$.
\end{thm}

\demthm \ref{infiniteanal}:\\
Every $y$ in $\mathcal{O}^{\C}$ is of the form
$$
y = g~D~g^{-1} - D = g \cdot 0
$$
for some $g \in G^{\C}$. Let us denote by $\mathfrak{k}$ the
commutator of $D$ in $\g$ and by $\mathfrak{m}$ its orthogonal.
For every $z = u\cdot 0$ in $\mathcal{O}$ (where $u \in G$), set
$\mathfrak{k}_{z} = \textrm{Ad}(u)(\mathfrak{k})$ and
$\mathfrak{m}_{z} = \textrm{Ad}(u)(\mathfrak{m})$. By Mostow's
Decomposition Theorem, $G^{\C}$ decomposes into the product
$G\!\cdot\!\exp(\ie~\mathfrak{m})\!\cdot\!\exp(\ie~\mathfrak{k})$.
Hence $y$ can be written as
$$
y  = u \exp(\ie~\b)\cdot 0 \qquad \b\in\mathfrak{m}, ~~u \in G
$$
or more conveniently as
$$
y = \exp(\ie~\a)\cdot z ,
$$
where $z = u\cdot 0$ belongs to $\mathcal{O}$ and $\a =
\textrm{Ad}(u)(\b)$ is an element in $\mathfrak{m}_{z}$. To show
that this expression of $z$ is unique and defines a projection
$\pi$ from $\mathcal{O}^{\C}$ onto $\mathcal{O}$ by $\pi(y) = z$,
it is sufficient to show that
\begin{equation}\label{mer}
\exp(\ie~\a)\cdot z \cap \g = \{z\}
\end{equation}
where $\a$ belongs to $\mathfrak{m}_{z}$(see Lemma
\ref{nonrecoupement} and Theorem \ref{projectionorbite}). By
$G$-equivariance, it is sufficient to show that the previous
equality for $z = 0$. For $\a \in \mathfrak{m}$, one has
$$
y = \exp(\ie~\a)\left(D\right)\exp(-\ie~\a) - D =
e^{\ie~\ad(\a)}\left(D\right) - D =
\cosh(\ie~\ad(\a))\left(D\right) - D +
\sinh(\ie~\ad(\a))\left(D\right),
$$
where $\cosh(\ie~\ad(\a))\left(D\right) - D$ belongs to $\g$ and
$\sinh(\ie~\ad(\a))\left(D\right)$ to $\ie\g$. Hence the condition
$\exp(\ie~\a)\cdot 0\in \g$ reads
$$ 0 = \sinh(\ie~\ad(\a))(D) = \ie\frac{\sinh(\ie~\ad(\a))}{\ie~\ad(
\a)}([\a, D]),$$ where $[\a, D]$ belongs to $\g$ since $D$ is a
derivation of $\g$. Since $\frac{\sinh(t)}{t}\geq 1$,  the
condition $ \sinh(\ie~\ad(\a))(D) = 0$ implies that $[\a, D]$
vanishes.  Thus $\a$ belongs to $\mathfrak{k}$. But $\a \in
\mathfrak{m}$ by hypothesis, so $\a = 0$. Consequently equality
\eqref{mer} is satisfied and the projection $\pi$ well-defined.
The definition of the homeomorphism $\Phi$ is as in the
finite-dimensional case. \cqfdt

\section{Stable manifolds}\label{section3}

In this section $G$ will denotes a semi-simple $L^{*}$-group of
compact type
 with Lie algebra $\g$ and
$G^{\C}$ the complex $L^{*}$-group with Lie algebra $\g^{\C} = \g
\oplus \ie\g$ characterized by the property that the natural
injection $\g \hookrightarrow \g^{\C}$ is the differential map of
an injection $G \hookrightarrow G^{\C}$. Recall that any compact
Lie group is an $L^{*}$-group, its Lie algebra being endowed with
the scalar product given by the Killing form. Let us suppose that
$G$ acts on a Banach weak K\"ahler manifold $(M, \omega, \gm, I)$
and preserves the K\"ahler structure of $M$. We use the following
convention. The Riemannian metric $\gm$, the symplectic form
$\omega$ and  the complex structure $I$ of $M$ are related by the
following formula~: $\omega(\cdot\,,\,\cdot) =
\gm(I\cdot\,,\,\cdot)$. Suppose  that the $G$-action extends to an
holomorphic action of $G^{\C}$ on $M$. Let $\mu~: M \rightarrow
\g'$ be a $G$-equivariant moment map of the $G$-action, where
$\g'$ denotes the continuous dual of $\g$. By definition~:
$$
d\mu_{x}(\a) = i_{X^{\a}}\omega,
$$
for $x\in M$, $\a \in \g$, where $X^{\a}$ denotes the Killing
vector field generated by the element $\a$ in $\g$. The following
Lemma is an extension of Lemma 7.2 page~96 in \cite{MFK}.

\begin{lem}\label{ghj}
Let $\xi\in \g'$ be  a $G$-invariant regular value of the moment
map $\mu$, and $x$ be an element of the level set
${\mu}^{-1}(\xi)$. Denote by $\mathfrak{k}_{x}$ the Lie algebra of
the isotropy group of $x$ and $\mathfrak{m}_{x}$ the orthogonal of
$\mathfrak{k}_{x}$ in $\g$. Then
$$
\exp(\ie \mathfrak{m}_{x})\cdot x~\cap~{\mu}^{-1}(\xi) = \{x\}.
$$
\end{lem}

\demlem \ref{ghj}:\\
Let $\a$ be an element in $\mathfrak{m}_{x}$ such that
$\exp(\ie\,\a)\cdot x$ belongs to the level set ${\mu}^{-1}(\xi)$.
Consider the real function $f$ defined by
$$
f(t) := \mu\left(\exp (\ie t\a)\cdot x \right)(\a).
$$
One has $f(0) = \mu(x)(\a) = \xi(\a)$ since $x$ belongs to the
level set ${\mu}^{-1}(\xi)$, and $f(1) = \mu\left(\exp(\ie
t\a)\cdot x\right)(\a)  = \xi(\a)$ by the hypothesis on $\a$. The
differentiability of $f$ implies that there exists $t_{0}$ in
$(0\,,\,1)$ such that the derivative of $f$ at $t_{0}$ vanishes.
One has
$$
0 = h'(t_{0}) = (d{\mu})_{y}\left(\ie \a\cdot y \right)(\a) =
i_{\a\cdot y }\omega(\ie \a\cdot y),
$$
where we have set $y := \exp(\ie t_{0}\a)\cdot x$. Since the
action of $G^{\C}$ is holomorphic, we have
$$0 = h'(t_{0}) =  i_{\a\cdot y}\omega\left(I
(\a\cdot y)\right)= \omega\left(\a\cdot y, I(\a\cdot y)\right) =
-\gm(\a\cdot y, \a\cdot y).
$$
It follows that $\a\cdot y = 0$. But
$$
\a\cdot y = \a\cdot \left(\exp(\ie t_{0}\a)\cdot x\right) =
\left(\exp(\ie t_{0}\a)\right)_{*}(\a\cdot x),
$$
hence $\a\cdot x$ vanishes also and $\a$ belongs to
$\mathfrak{k}_{x}$. But by hypothesis $\a$ belongs to
$\mathfrak{m}_{x}$, thus $\a$ vanishes. It follows that the
intersection of $\exp(\ie \mathfrak{m}_{x})\cdot x$ with the level
set is reduced to $\{x\}$. \cq

\begin{defe}
The stable manifold $M^{s}$ associated to the level set
${\mu}^{-1}(\xi)$ is defined as
$$
M^{s} := \{~ y \in M~|~ \exists~ g \in G^{\C},~ g\cdot y \in
{\mu}^{-1}(\xi) ~\}.
$$
\end{defe}

\begin{thm}\label{po}
There exists a $G$-equivariant projection from the stable manifold
associated to ${\mu}^{-1}(\xi)$ onto the level set
${\mu}^{-1}(\xi)$.
\end{thm}

\demthm \ref{po}:\\
Let us prove that every element in $M^{s}$ can be written uniquely
as
\begin{equation}\label{ecriture2}
y = \exp(\ie \a)\cdot z,
\end{equation}
where $z$ belongs to the level set ${\mu}^{-1}(\xi)$ and $\a$ to
the orthogonal $\mathfrak{m}_{z}$ in $\g$ of the isotropy algebra
$\mathfrak{k}_{z}$ of $z$. By definition of the stable manifold,
there exists $g$ in $G^{\C}$ and $x$ in ${\mu}^{-1}(\xi)$ such
that $y = g\cdot x$. By Mostow's Decomposition Theorem $G^{\C} =
G\!\cdot\!\exp(\ie \mathfrak{m}_{x})\!\cdot\!\exp(\ie
\mathfrak{k}_{x}),$ hence there exists $u \in G$,
$\mathfrak{b}\in\mathfrak{m}_{x}$ and $\mathfrak{c} \in
\mathfrak{k}_{x}$ such that $g = u \exp \ie \mathfrak{b} \exp \ie
\mathfrak{c}$. It follows that $y = (u \exp \ie \b)\cdot x$ since
$\exp \ie \mathfrak{k}_{x}$ acts trivially on $x$. But $u \exp \ie
\b = \exp\left(\ie \textrm{Ad}(u)(\b)\right) u$. Hence $y = \exp
(\ie \a)\cdot z$ where $z := u\cdot x$ belongs to  the level set
${\mu}^{-1}(\xi)$ and where $\a := \textrm{Ad}(u)(\b) \in
\mathfrak{m}_{z}$ since the scalar product is $G$-invariant. This
proves the existence of the expression \eqref{ecriture2}. In order
to proof uniqueness, let us suppose that
$$
\exp\left(\ie \a_{1}\right)\cdot z_{1} = \exp\left(\ie
\a_{2}\right)\cdot z_{2},
$$
for some $z_{1}$, $z_{2}$ in ${\mu}^{-1}(\xi)$, some $\a_{1}$ in
$\mathfrak{m}_{z_{1}}$ and some $\a_{2}$ in
$\mathfrak{m}_{z_{2}}$. One has~:
\begin{equation}\label{bla2}
\exp\left(-\ie \a_{2}\right)\exp\left(\ie \a_{1}\right)\cdot z_{1}
= z_{2}.
\end{equation}
By Mostow's Decomposition Theorem, there exists $u'$ in $G$,
$\a_{3}$ in $\textrm{m}_{z_{1}}$ and $\b_{3}$ in
$\mathfrak{k}_{z_{1}}$ such that
\begin{equation}\label{pdt2}
\exp\left(-\ie \a_{2}\right)\exp\left(\ie \a_{1}\right) =
u'\exp(\ie \a_{3})\exp(\ie \b_{3}).
\end{equation}
Thus equation \eqref{bla2} becomes
$$
u' \exp(\ie \a_{3})\cdot z_{1} = z_{2},
$$
or equivalently
$$
 \exp(\ie \a_{3})\cdot z_{1} = (u')^{-1}\cdot z_{2},
$$
But $(u')^{-1}\cdot z_{2}$ belongs to ${\mu}^{-1}(\xi)$ since
$z_{2}$ is an element of ${\mu}^{-1}(\xi)$ and $u' \in G$. Whence
Lemma \ref{ghj} implies that
$$
\exp(\ie \a_{3})\cdot z_{1} = (u')^{-1}\cdot z_{2} = z_{1}.
$$
It follows that $\a_{3}$ stabilizes $z_{1}$, hence vanishes
because it belongs to  $\mathfrak{k}_{z_{1}} \cap
\mathfrak{m}_{z_{1}} = \{0\}$, and that $z_{2} = u'\cdot z_{1}$.
Therefore equation \eqref{pdt2} becomes
$$
\exp\left(-\ie \a_{2}\right)\exp\left(\ie \a_{1}\right) =
u'\exp(\ie \b_{3}).
$$
It follows that
$$
\exp\left(\ie \a_{1}\right) = \exp\left(\ie
\a_{2}\right)u'\exp(\ie \b_{3}) = u'~\exp\left(\ie
\textrm{Ad}\left((u')^{-1}\right)(\a_{2})\right)~\exp(\ie \b_{3}),
$$
where  $\textrm{Ad}\left((u')^{-1}\right)(\a_{2})$ belongs to
$\textrm{Ad}\left((u')^{-1}\right)\left(\mathfrak{m}_{z_{2}}\right)
= \mathfrak{m}_{z_{1}}$. By uniqueness of Mostow's decomposition,
one  has
$$
u' = e, \qquad \a_{2} = \a_{1}, \qquad \textrm{and}~~\b_{3} = 0.
$$
Thus uniqueness of \eqref{ecriture2} is proved. The projection
$\pi$ is defined as follows~:
$$
\begin{array}{lcll}
\pi~:&   M^{s} & \twoheadrightarrow &
{\mu}^{-1}(\xi)\\
& y = \exp (\ie \a)\cdot z   & \mapsto & z,
\end{array}
$$
where in the expression of $y$, the element $z$ is supposed to
belong to the level set ${\mu}^{-1}(\xi)$ and $ \a$ to
$\mathfrak{m}_{z}$. The $G$-equivariance of the projection $\pi$
is a direct consequence of the identity
$$
u\cdot y = \exp \left(\ie \textrm{Ad}(u)(\a)\right)\cdot (u \cdot
z),
$$
and of the relation $\textrm{Ad}(u)(\mathfrak{m}_{z}) =
\mathfrak{m}_{u\cdot z}$. \cqfdt
\\
\\

\begin{rem}
{\rm In the case where $\mathcal{O}_{x}$ is a (irreducible)
Hermitian-symmetric coadjoint orbit of compact type, Theorem
\ref{projectionorbite} is a particular case of Theorem \ref{po} in
the following sense. The coadjoint action of $G^{\C}$ on
$\mathcal{O}_{x}^{\C}$ is holomorphic with respect to the natural
complex structure of $\mathcal{O}_{x}^{\C}$. The moment map
$\mu_{1}$ for the $G$-action on $\mathcal{O}_{x}^{\C}$
corresponding to the real symplectic form $\omega_{1}$ associated
to the natural complex structure of $\mathcal{O}_{x}^{\C}$ has
been computed in \cite{BG3}, p~153, formula~(3.14)~:
$$
\mu_{1}(y) = -i\left[ \frac{\pi(y)}{\kappa}, y \right],
$$
where $\kappa$ is a positive constant. It follows that the orbit
of compact type  $\mathcal{O}_{x}$ can be thought as the level set
${\mu_{1}}^{-1}(0)$ and $\mathcal{O}_{x}^{\C}$ as the associated
stable manifold. We believe that this picture is also faithful in
the case of a general orbit of compact type.}
\end{rem}

\!\!\hspace{-.5cm}\textbf{Acknowledgments.} In writing this work,
we benefit from the lecture given by P.~Gauduchon at the
University Paris~VII in 2000 to simplify some proves or use some
arguments (in particular the proof of Proposition \ref{log}
follows his argumentation). Therefore we are grateful to
P.~Gauduchon for his teaching and for useful discussions. Many
thanks also to T.~Ratiu for his kind hospitality at EPFL and for
providing a so nice working atmosphere. The interest of
T.~Wurzbacher for this subject is also acknowledged.

\end{document}